\documentclass{aa}  
\usepackage{graphicx}
\usepackage{longtable,lscape}
\begin{document}

   \title{A new wavelet-based approach for the automated treatment of large
sets of lunar occultation data\thanks{Algorithm tested with observations
collected at Calar Alto Observatory (Spain). Calar Alto is operated by the
German--Spanish Astronomical Center (CAHA).}}

   \titlerunning{A new wavelet-based approach for automated treatment of LO}

   \author{O. Fors
          \inst{1}\fnmsep\inst{2}
          \and
          A. Richichi\inst{3}
          \and
          X. Otazu\inst{4}\fnmsep\inst{5}
          \and
          J. N\'u\~nez\inst{1}\fnmsep\inst{2}
          }

   \offprints{O. Fors}

   \institute{Departament d'Astronomia i Meteorologia, Universitat de  
Barcelona, Mart\'{\i} i Franqu\'es 1, E-08028 Barcelona, Spain
              \email{ofors@am.ub.es}
         \and
             Observatori Fabra, Cam\'{\i} de l'Observatori s/n, E-08035 
Barcelona, Spain
         \and
             European Southern Observatory,
Karl-Schwarzschild-Str. 2, D-85748 Garching bei M\"unchen, Germany
             \email{arichich@eso.org}
         \and
             Computer Vision Center, Universitat Aut\`onoma de Barcelona, 
E-08193 Bellaterra, Spain
         \and
             Departament de Ci\`encies de la Computaci\'o, Universitat 
Aut\`onoma de Barcelona, E-08193 Bellaterra, Spain
             }

\date{Received November 4, 2007; accepted December 15, 2007}
 
  \abstract
   {The introduction of infrared arrays for lunar occultations (LO) work and the improvement of predictions based on new deep IR catalogues have resulted in a large increase in sensitivity and in the number of observable occultations.}
{We provide the means for an automated reduction of large sets of LO data. This frees the user from the tedious
task of estimating first-guess parameters for the fit of each LO lightcurve. At the end of the process, ready-made plots
and statistics enable the user to identify sources that appear to be resolved or binary, and to initiate their detailed interactive analysis.}
	 {
The pipeline is tailored to array data, including the extraction of the lightcurves from FITS cubes.
Because of its robustness and efficiency, the wavelet transform has been chosen 
to compute the initial guess of the parameters of the lightcurve fit.
}
   {We illustrate and discuss our automatic reduction pipeline by analyzing a large volume of novel
 occultation data recorded at Calar Alto Observatory.
The automated pipeline package is available from the authors.}
   {}

   \keywords{
Methods: data analysis --
Techniques: Image Processing --
Techniques: high angular resolution --
Astrometry --
Occultations
               }

   \maketitle
%
%________________________________________________________________
\section{Introduction}

For decades, lunar occultations (LO) have occupied a special niche as a
technique for high-angular resolution with excellent performance, but relatively
inefficient yield. The diffraction fringes that are created by the lunar limb as
it occults a background source, provide a unique opportunity to achieve
milliarcsecond angular resolution with single telescopes also of relatively
small diameter. In terms of instrumentation, LO have always been simple,
requiring only a fast photometer. Of course, they have the significant drawback
that only sources included in the apparent lunar orbit can be observed (about
10\% of the sky), and then only at arbitrary fixed times and with limited
opportunities for repeated observations. If one adds that each observation only
provides a one-dimensional scan of the source, it is clear that detailed and
repeated observations are better performed with long-baseline interferometry
(LBI), when available. One should, however, not forget additional important
advantages of LO: even for complicated sources, the full, one-dimensional
brightness profile can be recovered according to maximum-likelihood principles
without any assumptions on the source's geometry  (Richichi
\cite{1989A&A...226..366R}). Besides, the limiting sensitivity achieved in the
near-IR by LO at the 1.5\,m telescope on Calar Alto is K$\approx 8$\,mag
(Richichi et al.~\cite{2006A&A...445.1081R}). When extrapolated to a 4-meter
class telescope or larger, LO appear  quite competitive with even the most
powerful, LBI facilities (Richichi~\cite{1997IAUS..158...71R}).

As a result, although the trend is understandably to develop more flexible,
powerful and complex interferometric facilities, there is some balance that
makes LO still attractive at least for some applications. It should not be
forgotten that the majority of the hundreds of directly-measured stellar angular
diameters (Richichi~(\cite{vlti_2007}) listed 688, and the numbers keep
increasing) were indeed obtained by LO, and that LO are
still the major contributor to the discovery of small separation binary stars.

Two recent developments, however, have provided a significant boost to the 
performance of the LO technique, and have significantly enlarged its range 
of applications: a) the introduction of IR array detectors that can be 
read out at fast rates on a small subarray has made it possible to provide 
a large gain in limiting sensitivity, and b) IR survey catalogues
that have led to an exponential increase 
of the number of sources for which LO can be computed. Literally, 
thousands of occultations per night could now be potentially observed with 
a large telescope. We describe in this paper the details and impact of 
these two factors for LO work. We also address the new needs imposed on 
data reduction by the potential availability of a large volume of lunar 
occultation data per night, by describing new approaches to an automated 
LO data pipeline. We illustrate both the new quality of LO data and their 
analysis by means of examples drawn from the observation of two recent 
passages of the Moon over crowded regions in the vicinity of the Galactic 
Center, carried out with array-equipped instruments at Calar Alto and 
Paranal observatories.

%__________________________________________________________________
\section{Infrared arrays and new catalogues}

A number of reasons make the near-IR domain preferable for LO work with respect to other wavelengths.

First, LO observations are affected by the high background around the Moon
which, being mainly reflected solar light, shows an intensity maximum at visible
wavelengths. Because of the atmospheric Rayleigh scattering
($\propto\lambda^{-4}$), the background level greatly decreases in the near-IR. 
At longer wavelengths ($10{\mu}{\rm} m-20{\mu}{\rm m}$), the thermal emission
of  Earth's atmosphere and of the lunar surface introduces a high-background level.

Second, the spacing of diffraction fringes at the telescope is proportional to
$\lambda^{-\frac{1}{2}}$. Therefore, for two LO observations with the same
temporal sampling, one recorded in IR will obtain a higher fringe sampling  than
one in the visible.

Finally, at least in the field of stellar diameters, there is an advantage to
observing in the near-IR because for a given bolometric flux redder stars
will present a larger angular diameter.

Being cheap and with a fast time response,
near-IR photometers have traditionally represented the detector
of choice for LO observations.
Richichi~(\cite{1997IAUS..158...71R}) showed the
great increase in sensitivity possible with
panoramic arrays, which by reading only the pixels of interest,
permit to avoid most of the shot noise generated by the
high background in LO.
Such arrays are now becoming a viable option,
thanks to  read-out noises, that are decreasing at each
new generation of chips, and to flexible electronics allow us to address a subarray and read it out at millisecond rates.
Richichi~(\cite{1997IAUS..158...71R}) predicted that an 8\,m telescope
would reach between K=12 and 14\,mag, depending on the
lunar phase and background, with an integration time of 12\,ms at
signal--to--noise ratio (SNR)=10. Observations on one of the 8.2\,m VLT telescopes, equipped
with the ISAAC instrument in the so-called burst mode
(Richichi et al. \cite{2006Msngr.126....24R}), show a limiting
magnitude K$\approx$12.5 at SNR=1 and 3ms integration time,
in agreement with the decade-old prediction.

These newly-achieved sensitivities call for a corresponding extension in the
limiting magnitudes of the catalogues used for LO predictions, and their
completeness.
In the near-IR, until recently the only survey-type catalogue
available was the Two-Micron Sky Survey (TMSS, or IRC, Neugebauer \&
Leighton~\cite{1969tmss.book.....N}) that was incomplete in declination and
limited  to K$<3$.  Already, a 1\,m-class telescope equipped with
an IR photometer exceeds this sensitivity by several magnitudes
(Fors et al.~\cite{2004A&A...419..285F}, Richichi et al.
\cite{1996AJ....112.2786R}).
The release of catalogues associated with modern all-sky near-infrared surveys, such as 2MASS (Cutri et
al.~\cite{2003yCat.2246....0C}) and DENIS (Epchtein et al.~\cite{1997Msngr..87...27E}), has helped.
Our prediction software {\tt ALOP} (Richichi~\cite{AR_thesis}) includes about 50
other catalogues with stellar and extragalactic sources.
We have now added a subset of 2MASS
with K$\le11$, which includes $3.7\times10^6$ sources subject to
occultations. 

\begin{figure*}
\centering
\includegraphics[width=18cm]{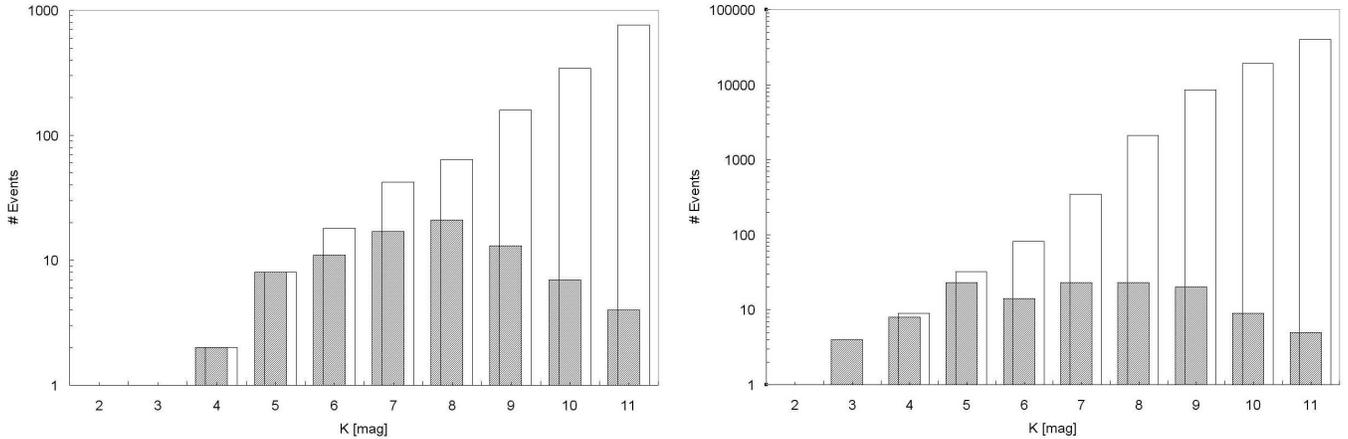}
\caption{Frequency of lunar occultation events as a function
of K magnitude, computed on the basis
of all standard catalogues in {\tt ALOP} (gray bars) and of the
2MASS catalogue only (limited to $K\le11$, clear bars).
For both cases, we have used the constraints
of Moon $\ge 25\degr$ above horizon and Sun $\le -5\degr$ below horizon.
Left: a relatively rich 5-night run, from
7 thru 11 January 2006, at Calar Alto Observatory.
Right: part of the night of August 5, 2006 from Paranal, when the
Moon reached a minimum approach of $12^{\prime}$ from the Galactic Center.
Note the logarithmic scale.
}
\label{fig:lo_alop}
\end{figure*}

While with the previous catalogues a typical night run
close to the maximum lunar phase would cover 100-150 sources over several nights
, predictions with 2MASS can include
thousands of events observable with a large telescope over one night.
Special cases, like the passage of the Moon over crowded, obscured regions
in the direction of the Galactic Center, can include 
thousands of events predicted over just a few hours
(Richichi et al. \cite{2006Msngr.126....24R}, Fors et al. \cite{SEA_2006}).
Fig.~\ref{fig:lo_alop} illustrates the two cases. 
The incompleteness of the catalogues without 2MASS is evident already
from the regime $5\le {\rm K}\le 7$\,mag. At even fainter magnitudes, but
still within the limits of the technique as described here,
the predictions based  on the 2MASS catalogue are more numerous by
several orders of magnitude.

%------------------------------------------------------------------------------
%------------------------------------------------------------------------------

Note that the increase in the number of potential
occultation candidates is not reflected automatically in more
results. The shift to fainter
magnitudes implies that the SNR of the recorded lightcurves
is on average lower; 
LO runs based on 2MASS predictions are now likely to be
less efficient in detecting binaries when compared, for example, to 
studies such as those of 
Evans et al.~(\cite{1986AJ.....92.1210E}) and
Richichi et al.~(\cite{2002A&A...382..178R}),
especially for large brightness ratios.

%__________________________________________________________________
\section{Automated reduction of large sets of lunar occultation data}

In general, LO data are analyzed by fitting model lightcurves. We take as an
example the Arcetri Lunar Occultation Reduction software ({\tt ALOR}), a general
model-dependent lightcurve fitting algorithm first developed by one of us
(Richichi \cite{1989A&A...226..366R}). Two groups of parameters are
simultaneously fitted using a non-linear least squares method. First, those
related to the geometry of the event: the occultation time ($t_0$), the stellar
intensity ($F_0$), the intensity of the background ($B_0$) and the limb linear
velocity with respect to the source ($V_{\rm P}$). Second, those related to
physical quantities of the source: for resolved sources; the angular diameter
and; for binary (or multiple) stars, the projected separation and the brightness
ratio of the components. 

In general, the fitting procedure is approached in two steps. First, a
preliminar fit assuming an unresolved source model is performed. To ensure convergence, {\tt ALOR} needs to be provided with reliable initial guesses.
We can estimate the geometrical parameters with a visual inspection of the data, and $V_{\rm P}$ is predicted. 
The source parameters can be refined in a second step.
This is done interactively since it requires understanding the nature of
each particular lightcurve and the possible correlation between geometrical and
physical parameters. 

\begin{figure*}
\centering
\includegraphics[height=18cm]{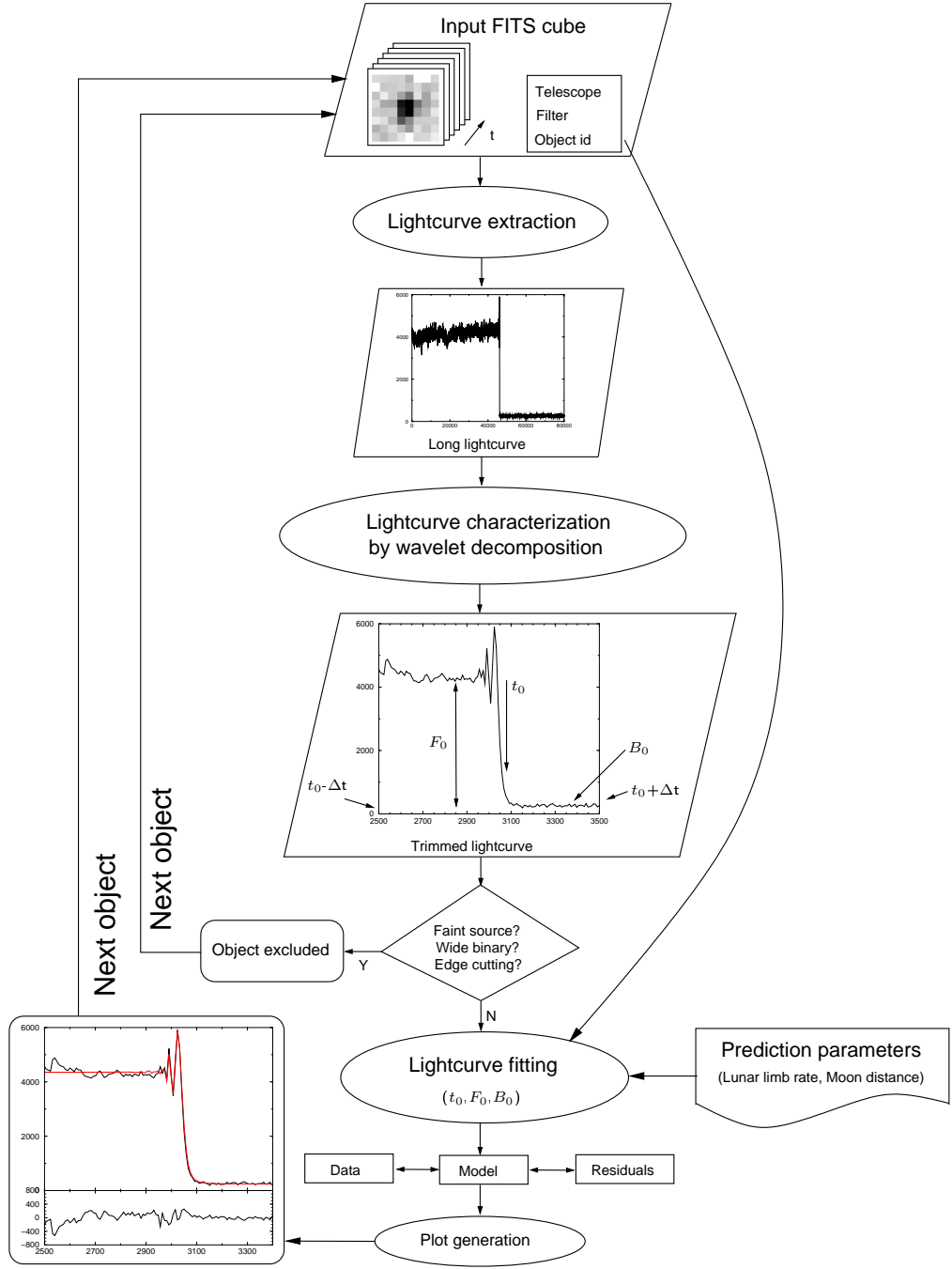}
\caption{Flow-chart description of {\tt AWLORP}.}
\label{fig:lo_pipeline}
\end{figure*}

As a result of that great increase in the number of potential occultations,  we
soon realized that we needed a substantial optimization in the processes of
extracting the occultation lightcurves from the raw data and of the interactive
evaluation of the LO lightcurves for the estimate of the initial parameter
values needed for the fits. We then developed, implemented, and tested a new
automatic reduction tool, the Automatic Wavelet-based Lunar Occultation
Reduction Package ({\tt AWLORP}; (Fors~\cite{2006PhDT.........9F}). This allows
both lightcurve extraction and characterization to perform the preliminary
analysis of large sets of LO events in a quick and automated fashion. In the
following, we describe the main parts of {\tt AWLORP}, which are schematically
illustrated in Fig.~\ref{fig:lo_pipeline}.
%------------------------------------------------------------------------------
%------------------------------------------------------------------------------

%___________________________________________
\subsection{Input data and lightcurve extraction}

In the cases available to us, the LO data are stored in Flexible Image Transport
System (FITS) cubes. The number of cube frames is given by the frame exposure
and total integration time. Additional information, such as telescope diameter,
filter and identificator of the occulted object, are extracted from the  FITS
cube header and saved in a separate file. In addition, the limb linear velocity
and the distance to the Moon as predicted by {\tt ALOP} are available in a
separate file.

An occultation lightcurve must be extracted from the recorded FITS cube file.
We explored several methods for this purpose, among them
fixed aperture integration, border clipping, Gaussian profile and brightest-faintest
pixels extraction. We found these partly unsatisfactory, among
other things, because of lack of connectivity across the stellar image
and because of sensitivity to flux and image shape variations.

We addressed the problem of connectivity with the use of
masking extraction, and two methods were
considered.  The first method, called {\tt 3D-SExtractor}, consists of a
customization of the object detection package {\tt SExtractor} (Bertin \&
Arnouts~\cite{1996A&AS..117..393B}) for the case of 3D FITS LO cubes. The
algorithm invokes {\tt SExtractor} for every frame and evaluates its output to
decide if the source has been effectively detected. The segmentation map (or
source mask) provided by  {\tt SExtractor} defines the object (background)
pixels in case of positive (negative) detection. These pixels are used to
compute the source (background) intensity before and after the occultation. The
second method, called {\tt Average mask}, consists in performing simple aperture
photometry using a predefined source mask. This is obtained by averaging a large
number of frames previous to the occultation and by applying a 3$\sigma$
thresholding.

We empirically compared {\tt 3D-SExtractor} and {\tt Average mask} methods under
a variety of SNR, scintillation, and pixel sampling situations. Although the
{\tt 3D-SExtractor} makes use of a more exact mask definition for every frame,
{\tt Average  mask} was found to provide less noisy lightcurves with no evident
fringe smoothing. Therefore, we adopted this extraction algorithm as the default in the {\tt AWLORP} description.

%___________________________________________
\subsection{Lightcurve characterization}

Inaccuracies in catalogue coordinates and lunar limb irregularities introduce an
uncertainty in the predicted occultation time of about 5 to 10 seconds. To secure the effective registering of an occultation event, the acquisition
sequence is started well before the predicted occultation time. This results in
a very long extracted lightcurve, typically spanning several tens of seconds. In
contrast, the fringes that contain the relevant high-resolution information
extend only a few tenths of a second. In addition, to accomplish a proper
fitting of this much shorter lightcurve subsample, as mentioned before, we need reliable estimates of $t_{0}$, $B_{0}$ $F_{0}$. 

The problem corresponds to detecting a slope with a known-frequency range in a
noisy, equally sampled data series. The key idea here is to note that the drop
from the first fringe intensity (close to $t_{0}$) is always characterized by a
signature of a given spatial frequency. Of course, this frequency depends on the
data sampling but, once this is fixed, the aimed algorithm should be able to
detect that signature and provide an estimate of $t_{0}$, regardless its SNR.
Once $t_{0}$ is known, the other two parameters ($B_{0}$ and $F_{0}$) can be
estimated.

This problem calls for a transformation of the data that would be capable of
isolating signatures in frequency space, while simultaneously keeping the
temporal information untouched. Wavelet transform turns out to be convenient for
this purpose.

%%%%%%%%%%%%%%%%%%%%%%%%%%%%%%%%%%%%%%
\subsubsection{Wavelet transform overview}
\label{subsubsect:wav_over}

The wavelet transform of a distribution $f(t)$ can be expressed as:
%------------------------------------------------------------------------------
\begin{eqnarray}
\label{WT_cont}
W(f(t))(a,b) = \vert a\vert^{-{1 \over 2}}\int_{-\infty}^{+\infty} f(t)
\,\psi\biggl({t-b\over a}\biggr)\;dt\;,
\end{eqnarray}
%------------------------------------------------------------------------------
\noindent where $a$ and $b$ are scaling and translational parameters
respectively. Each base (or {\it scaling}) function $\psi({t-b\over a})$ is a scaled and translated version of a function $\psi$ called {\it mother wavelet}, satisfying the relation $\int \psi({t-b\over a})=0$.

We followed the {\it \`{a} trous} algorithm (Starck \&
Murtagh~\cite{1994A&A...288..342S}) to obtain the discrete wavelet decomposition
of $f(t)$ into a sequence of approximations:
%------------------------------------------------------------------------------
\begin{equation}
F_1(f(t)) = f_1(t),\;\;
F_2(f_1(t)) = f_2(t),\cdots\;.
\end{equation}
%------------------------------------------------------------------------------

$f_i(t)\;(i=1,\cdots,n)$ are computed by performing successive convolutions with
a filter derived from the {\it scaling} function, which in this case is a $B_3$
cubic spline. The use of a $B_3$ cubic spline leads to a convolution with a mask
of 5 elements, scaled as {\tt (1,4,6,4,1)}.

The differences between two consecutive approximations $f_{i-1}(t)$ and $f_i(t)$
are the wavelet (or {\it detail}) planes, $w_i(t)$. Letting $f_0(t)=f(t)$, we
can reconstruct the original signal from the expression: 
%------------------------------------------------------------------------------
\begin{equation} 
f(t) = \sum_{i=1}^{n}w_i(t) + f_r(t)\;\;, 
\label{wav_des}
\end{equation}
%------------------------------------------------------------------------------
where $f_r(t)$ is a residual signal that contains the global energy of $f(t)$.
Note that $n=r$, but we explicitly substitute $n$ with $r$ to clearly
express the concept of $residual$. Each wavelet plane can be understood as a
localized frequential representation at a given scale according to the
wavelet base function used in the decomposition.  

In our case, we are using a multiresolution decomposition scheme, which means the original signal $f(t)$ has twice the resolution of $f_1(t)$. This latter has twice the resolution of $f_2(t)$, and so on.

%%%%%%%%%%%%%%%%%%%%%%%%%%%%%%%%%%%%%%
\subsubsection{Algorithm description} 
\label{subsubsect:alg_des}

We developed a program to perform a discrete decomposition of the lightcurve
into $n_{\rm wav}$ wavelet planes. Note that the choice of $n_{\rm wav}$ depends
exclusively on the data sampling and will be discussed later.
For example, $n_{\rm wav}=7$ was empirically
found to be a suitable value for representing all the features in the frequency
space of the lightcurve when the sampling was $8.4$~ms. The 2nd to 7th wavelet
planes resulting from the decomposition of the lightcurve of the bright star
\object{SAO~190556} (SNR=43) are represented in
Fig.~\ref{fig:wavelet_estimation}. The 1st plane was excluded as it nearly
exclusively contains noise features not relevant for this discussion. For the
sake of simplicity, we will consider this particular lightcurve and sampling
value in the description that follows.

We designed an algorithm which estimates $t_{0}$, $B_{0}$ and $F_{0}$ from the
previous wavelet planes. This consists of the following two steps: First, it was
empirically determined\footnote{This was realized by repeating the same analysis
to many other lightcurves of different SNR values and same time sampling
($8.4$~ms).} that the 7th plane serves as an invariant indicator of the
occultation time ($t_{0}$). In particular, $t_{0}$ coincides approximately
with the zero located between the absolute minimum ($t_{0}^{min}$) and maximum
($t_{0}^{max}$) of that plane (see upper right panel in
Fig.~\ref{fig:wavelet_estimation} for a zoomed display of the 7th plane). The
good localization of $t_{0}$ in this plane is justified because the first fringe
magnitude drop is mostly represented at this wavelet scale. In addition, the
presence of noise is greatly diminished in this plane. This is because noise
sources (electronics or scintillation) contribute at higher frequencies, and
therefore are better represented at lower wavelet scales (planes). In other
words, this criteria for estimating $t_{0}$ is likely to be insensitive to
noise, even for the lowest SNR cases. 

Second, once a first estimate of $t_{0}$ was obtained, $B_{0}$ and $F_{0}$ could
be derived by considering the 5th wavelet plane. We found that this plane
indicates those values with fairly good approximation. The procedure is
illustrated in Fig.~\ref{fig:wavelet_estimation} and is described as follows:

%------------------------------------------------------------------------------
\begin{figure*}
\centering
\resizebox{\hsize}{!}{\includegraphics{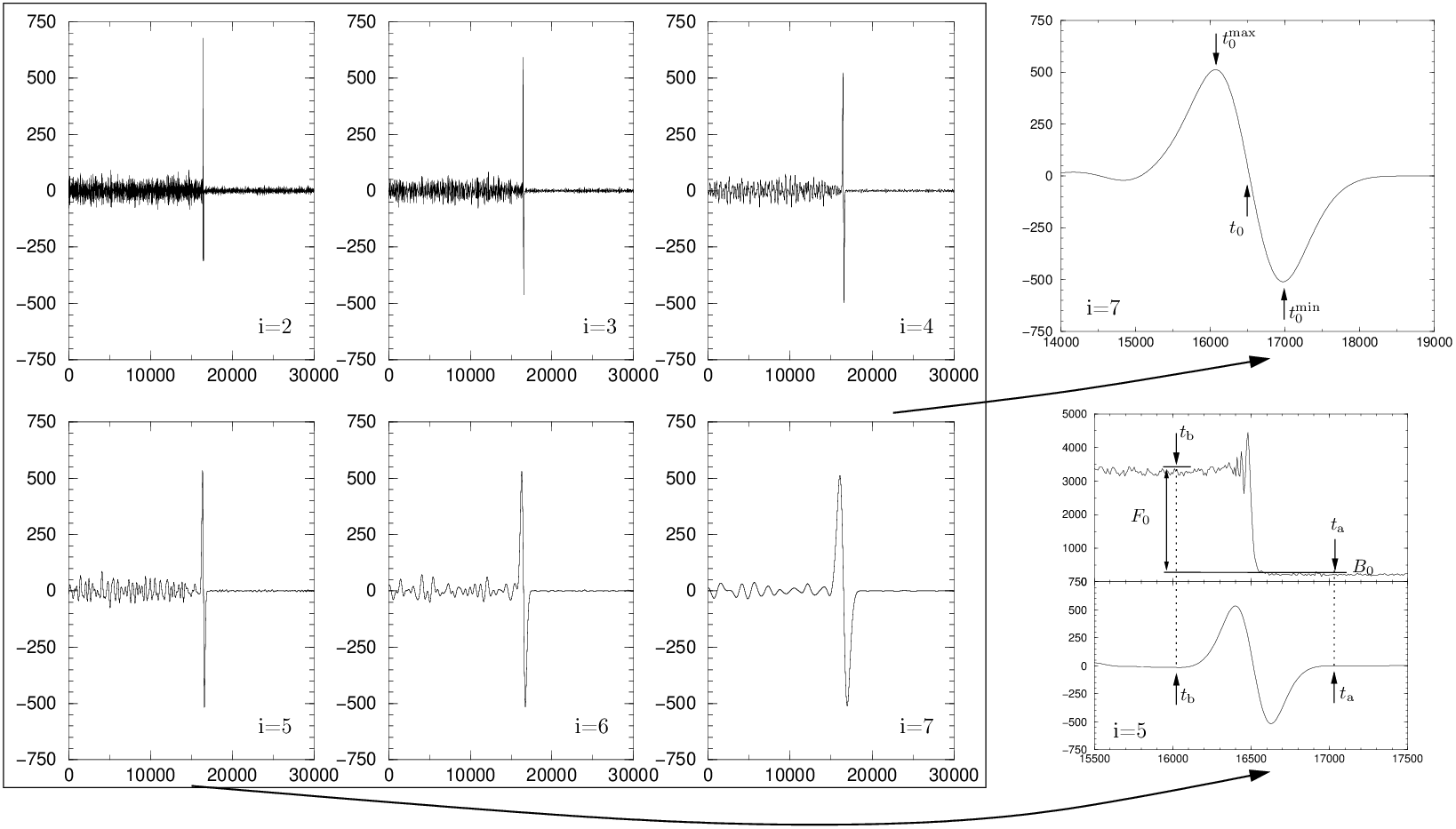}}
\caption{Schematic of the wavelet-based algorithm for the
estimation of $t_{0}$, $F_{0}$, and $B_{0}$ to be used in {\tt
AWLORP}. The lightcurve corresponds to an occultation of \object{SAO~190556}
observed at the Calar Alto Observatory, sampled every $8.4$~ms. Left: box
with 2nd to 7th wavelet planes resulting from the wavelet decomposition of the
original lightcurve. Upper right: the 7th plane is found to be a good indicator of
$t_{0}$. A zoomed display of the region around $t_{0}$ is shown. Lower right: a
box display of 5th plane (bottom part of this panel) provides the abscissae $t_{\rm b},t_{\rm a}$ to compute $F_{0}$ and $B_{0}$ in the
original lightcurve (upper part of the same panel).}
\label{fig:wavelet_estimation}
\end{figure*}
%------------------------------------------------------------------------------

\begin{enumerate}

\item We consider the abscissa in the 5th plane, corresponding to
$t_{0}$ found in the 7th plane.

\item From $t_{0}$, we search for the nearby zeroes in the 5th plane, before and after the above abscissa.  We call them $t_{\rm b}$ and $t_{\rm a}$.

\item We estimate $B_{0}$ by averaging the lightcurve values around $t_{\rm
a}$ within a specified time range. We empirically fixed this to
$\left[-8,8\right]$ samples because it provided a good compromise between
improving noise attenuation and suffering from occasional background slopes.

\item The same window average is computed around $t_{\rm b}$. The obtained
value ($I_{\rm p}$) represents a mean value of the intensity at the {\it
plateau} region before the onset of diffraction fringes. Note that the 5th wavelet
plane was chosen because its zero at $t_{\rm b}$ is safely before the
fringes region in the lightcurve, where the intensity is not constant and, thus,
not appropriate for $I_{\rm p}$ calculation.

\item $F_{0}$ is computed by subtracting $B_{0}$ to $I_{\rm p}$.

\end{enumerate}

As in the case of the 7th plane, the contribution in the 5th plane is dominated
by signal features represented at this scale, while noise, even the
scintillation component, has a minor presence. Therefore, again, the estimation
criteria for $B_{0}$ and $F_{0}$ is likely to be well behaved and robust in
presence of high noise.

Although {\tt AWLORP} was demonstrated on a particular data set,
its applicability is totally extensible to any sampling of the
lightcurve and also to reappearances. To show this,
we repeated the previous algorithm description for 6 sets of 100
simulated\footnote{The procedure folowed to simulate these data sets is
explained in Sect.~\ref{subsect:simul_data}.} lightcurves of different samplings
(1,2,4,6,8 and 10\,ms). For these six samplings, $n_{\rm wav}$ was found to be 8,7,6,6,5 and 5, respectively. Note these values are proportional to a geometric sequence of ratio 2 and argument $(8-n_{\rm wav})$, which is in agreement with the dyadic nature of the wavelet transform we adopted.

%___________________________________________
\subsection{Lightcurve fitting}

The algorithm just described has been integrated in
an automated pipeline. As shown in the scheme of
Fig.~\ref{fig:lo_pipeline}, the characterization of the lightcurve  is used to
decide if  a fit can be performed succesfully with {\tt ALOR}. The cases of very
faint sources, wide binaries and those lightcurves with some data truncation
(i.e. very short time span on either side of the diffraction fringes) are the
typical exclusions, and are discussed in Sect.~\ref{problems}.
In case of positive evaluation, {\tt ALOR} is executed using
the detected values of $t_0$,$F_0$, and $B_0$ as initial guesses. After the
preliminary fit is performed, a quicklook plot of lightcurve data, model, and
residual files is generated. This process is iterated for all the observed
sources.

This automatic pipeline frees us from the most tedious and error-prone part
of {\tt ALOR} reduction. The pipeline spends a few seconds
per occultation to complete the whole process described in
Fig.~\ref{fig:lo_pipeline}. For comparison, an experienced user takes 10-20
minutes per event for reaching the same stage of the reduction
pipeline. In cases when the data sets included hundreds of occultation events,
this difference is substantial.
The pipeline was coded entirely in {\tt Perl} programming language, which turns
our to be a powerful and flexible tool for concatenating the I/O streams of
independent programs.

Once {\tt AWLORP} has automatically generated all the single source fit plots,
the user can perform a quick visual inspection.  The
objective of this first evaluation is to separate the unresolved, relatively
uninteresting events from those that bear the typical marks of a resolved
angular diameter, of an extended component or of a multiple source.
These latter will still need an interactive data reduction with
{\tt ALOR}, but they will represent typically only a small fraction of the whole
data set.

%__________________________________________________________________
\section{Performance evaluation}

We have verified the performance of {\tt AWLORP} by analysing both simulated and
real LO data sets.

%___________________________________________
\subsection{Simulated data}
\label{subsect:simul_data}

%------------------------------------------------------------------------------
\begin{figure*}
\centering
\includegraphics[height=10cm]{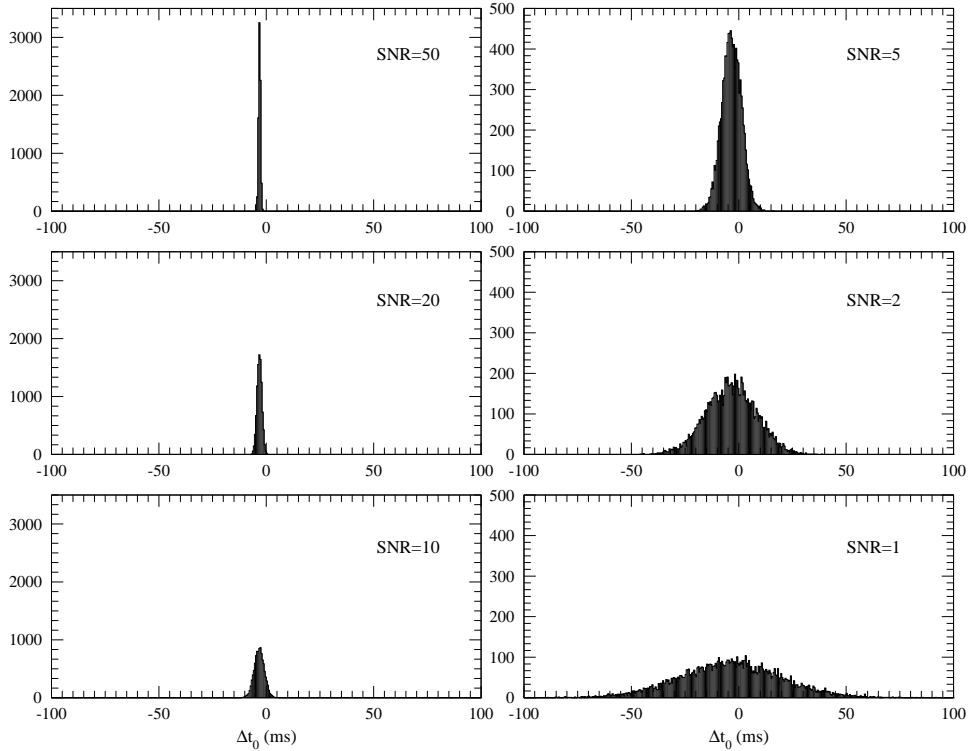}
\caption{Application of {\tt AWLORP} to six sets of 10000 simulated lightcurves
at 2ms sampling and of different SNRs values. As explained in the
text, the offset between the simulated
occultation time and the time detected by {\tt AWLORP} (${\Delta}t_{0}$) is
Gaussian distributed with FWHMs inversely proportional to the SNR value, and
the histogram peaks are sistematically shifted within the range
${\Delta}t_{0}\sim\left[-4,-2\right]$ms (only 1 to 2 sampling points).}
\label{fig:snr_simulated}
\end{figure*}
%------------------------------------------------------------------------------

Thanks to a specific module included in {\tt ALOR}, a set of simulated LO
lightcurves was generated for varying SNR values. The noise model assumes three
independent noises sources: detector electronics, photon shot-noise, and
scintillation, which are of Gaussian, Poisson, and multiplicative nature,
respectively (Richichi \cite{1989A&A...226..366R}).
With a realistic combination of these three noise sources, we generated six 
series with SNR $50,20,10,5,2$ and $1$, each of them
consisting of 10000 lightcurves. We chose the sampling to be 2ms, which is a realistic value considering what is offered by current detectors.

{\tt AWLORP} was executed for all the 60000 simulated events. For each
lightcurve, we found an estimate of the triplet ($t_0$,$F_0$,$B_0$). The {\tt
AWLORP} only failed to characterize the lightcurve in 10 cases of the noisiest
series for which the {\tt ALOR} fits could not converge. For the remaining
59990 cases, we computed the difference (${\Delta}t_{0}$) between the detected and the simulated occultation time and plotted these differences as shown in
Fig.~\ref{fig:snr_simulated}. Two comments can be made. 

First, the ${\Delta}t_{0}$ distribution is, to a good approximation,
Gaussian-shaped.
This is in agreement with the fact that the first fringe
signature is primarily dominated by Gaussian noise at the wavelet plane
($n_{wav}=7$) employed to estimate $t_0$. This noise distribution has its
origins in the detector read-out for the faint end (low SNR) and in the shot-noise for the bright end (high SNR), which can be approximated by a Gaussian distribution in this regime.
In addition, the typical width of the ${\Delta}t_{0}$ distribution is inversely
proportional to the SNR value. A gaussian function was fitted to every histogram, and we found the values $\sigma =23.0,11.7,4.6,2.3,1.1,0.5$ for the cases with SNR$=1,2,5,10,20,50$.

Second, note that the histograms in Fig.~\ref{fig:snr_simulated} are not exactly
centered at ${\Delta}t_{0}=0$, but systematically shifted $4$ms to $2$ms
(only $2$ to $1$ sampling points). This error is about the Nyquist cut-off
frequency of our data sampling. It can be assumed as a limitation imposed
by the data and not as an intrinsic constraint of {\tt AWLORP}. 
The difference could be corrected by subtracting this small offset to 
all analyzed lightcurves, but it is in any case of no consequence
for the purpose of the subsequent interactive analysis.

%___________________________________________
\subsection{Real data}

We considered a set of six real lightcurves. These were recorded in the course
of Calar Alto Lunar Occultation Program (CALOP) (Richichi et
al.~\cite{2006A&A...445.1081R}, Fors et al.~\cite{2004A&A...419..285F}). They
correspond to a series of SNR values similar to the one discussed in
Sect.~\ref{subsect:simul_data}. 

%------------------------------------------------------------------------------
\begin{figure*}
\centering
\includegraphics[height=18cm]{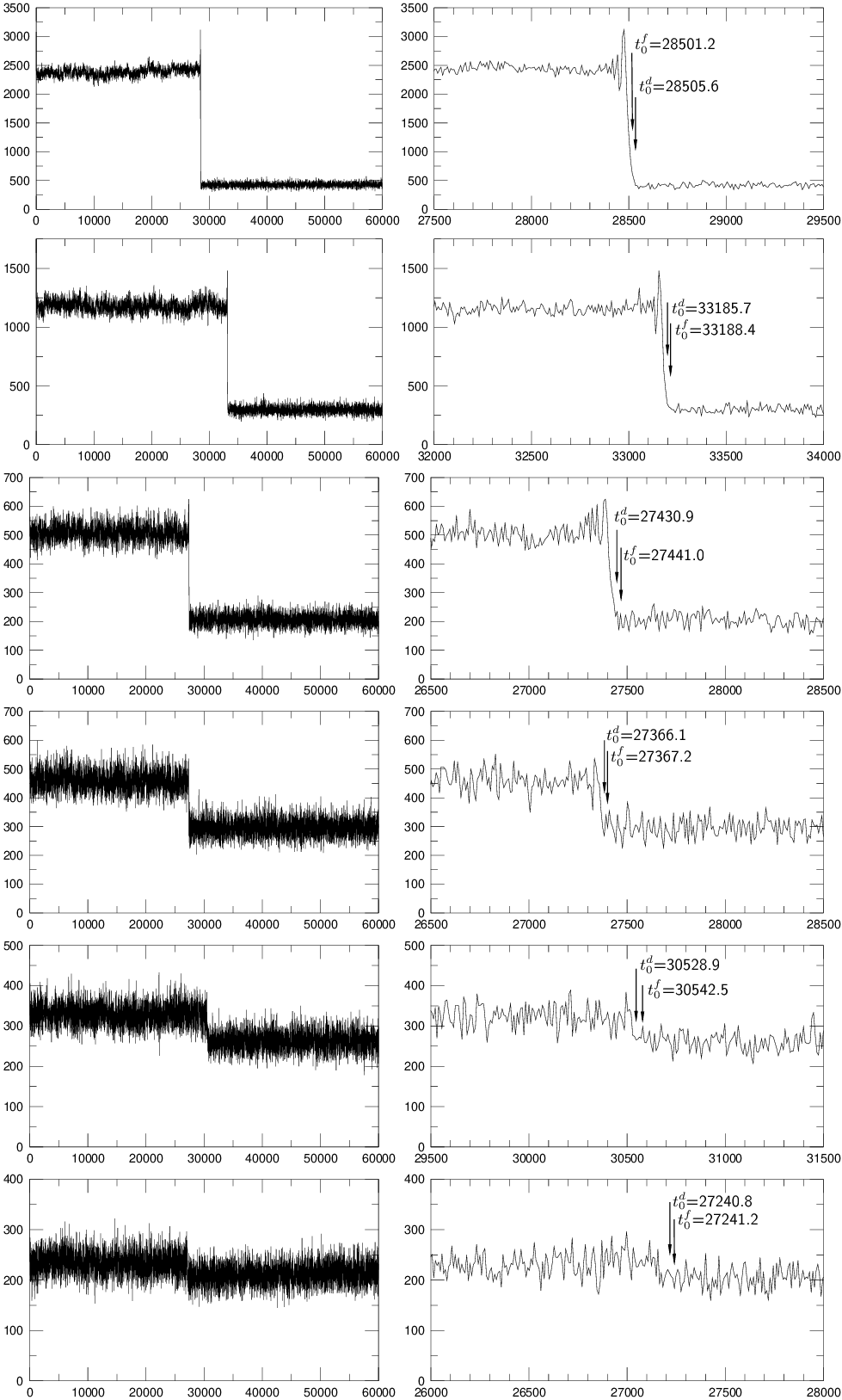}
\caption{Application of {\tt AWLORP} to 6 lightcurves with different SNRs (from
top to bottom: 47.2, 22.3, 10.9, 5.9, 2.1 and 1.2) observed as part of the CALOP
program. The left side panels show the whole lightcurves 
(60 seconds). The right side panels show the trimmed lightcurves (spanning
only 2 seconds) around the $t_{0}^d$ value detected by {\tt AWLORP}. The
occultation time fitted by {\tt ALOR} using $t_{0}^d$ as initial value,
$t_{0}^f$, are also displayed. Note that even in the faintest SNR case, the
occultation time is correctly detected.}
\label{fig:wavelet_estimation_snr}
\end{figure*}
%------------------------------------------------------------------------------

The robustness of $t_{0}$ estimation is shown in
Fig.~\ref{fig:wavelet_estimation_snr}, where even in the lightcurves at the
limit of detection (SNR$=1.2, 2.1$) the value of $t_{0}$ is correctly detected.
This is confirmed by visual inspection and by 
an comparison with the predicted values.

To verify this concordance, we ran {\tt ALOR} fits for all six lightcurves
with the {\tt AWLORP}-detected triplets ($t_0$,$F_0$,$B_0$) as initial values.
Even in the faintest cases, {\tt ALOR} converged for all parameters of the
lightcurve model. With regard as $t_{0}$, the difference between the initial and
the fitted values never exceeded $13.6$~ms ($1.6$ sample points) as can be seen
in Fig.~\ref{fig:wavelet_estimation_snr}.

\subsection{Problematic cases}\label{problems}
The pipeline just described works well for about 98\% of the recorded events. There
are, however, a few special situations where the algorithm of
Fig.~\ref{fig:lo_pipeline} fails. Those can be classified in three distinctive
groups:

\begin{enumerate}

\item The current version of wavelet-based lightcurve characterization does not
support wide binary events. In other words, the pipeline cannot simultaneously
determine the values ($t_{0}^{A}$, $B_{0}^{A}$ and $F_{0}^{A}$) and
($t_{0}^{B}$, $B_{0}^{B}$ and $F_{0}^{B}$) for two components $A$ and $B$
separated by more than a hundred of milliarcseconds. 
Since these cases represent at most a few percent 
of the overall volume of LO events and they are also relatively
uninteresting, this feature has not been implemented yet.

\item Due to observational constraints, to
an unusually large prediction 
error or simply by mistake, sometimes 
the recording of an event is started too close
to the actual occultation time.  
Since the scaling
function has a given size at each wavelet scale, there is a filter ramp
that extends over an initial span of data depending on the wavelet plane.
For example, in the case of data in
Sect.~\ref{fig:wavelet_estimation_snr} this happens up to 4000 milliseconds from
the beginning of the lightcurves, since this is the size of the scaling function
at the scale of the 7th plane for the given temporal sampling.

\item  Depending on the subarray size employed, the image scale, the seeing
conditions or telescope tracking, part of the stellar image might be displaced
outside the subarray so that the extracted flux decreases and the shape of
lightcurve is affected. Under these circumstances, {\tt AWLORP} is likely to
produce false $t_{0}$ detections. Again, the small number of cases affected
does not justify the substantial effort required to improve the {\tt AWLORP} treatment.

\end{enumerate}

%______________________________________________________________
\section{Summary}
The observation of lunar occultation (LO) events with modern
infrared array detectors at large telescopes, combined with the
use of infrared survey catalogues for the predictions, has shown
that even a few hours of observation can result in many tens if
not hundreds of recorded occultation lightcurves. The work
to bring these data sets to a stage where an experienced observer
can concentrate on accurate interactive data analysis for the
most interesting events is long and tedious.

We have designed, implemented, and tested an automated data
pipeline that takes care of extracting the lightcurves from the
original array data (FITS cubes in our case); of restricting the
range from the original tens of seconds to the few seconds 
of interest near the occultation event; of estimating the initial
guesses for a model-dependent fit; of performing the fit; and
finally of producing compact plots for easy visual inspection.
This effectively reduces the time needed for the initial
preprocessing from several days to a few hours, and frees
the user from a rather tedious and error-prone task.
The pipeline is based on an algorithm for automated extraction
of the lightcurves,
and on a wavelet-based algorithm for the estimation of the
initial parameter guesses.

The pipeline has been tested on a large number 
of simulated lightcurves spanning a wide range of realistic
signal-to-noise ratios.
The result has been completely satisfactory: in all
cases in which the algorithm converged, the derived lightcurve
characterization was correct and consistent with the simulated
values. Convergence could not be reached due to poor signal-to-noise ratio in only in ten cases out of 60000. These cases would, of
course, be challenging for an interactive data analysis
by an experienced observer as well. We also tested the pipeline on
a set of real data, with similar conclusions. 
We identified and discussed the cases that may prove
problematic for our scheme of automated preprocessing.

\begin{acknowledgements}
This work is partially supported by the ESO 
Director General's Discretionary Fund
and by the
\emph{MCYT-SEPCYT Plan Nacional I+D+I AYA \#2005-082604}.
\end{acknowledgements}


\begin{thebibliography}{}

\bibitem[1996]{1996A&AS..117..393B} Bertin, E., \& 
Arnouts, S.\ 1996, \aaps, 117, 393 

\bibitem[2003]{2003yCat.2246....0C} Cutri, R.~M., Skrutskie, M.~F., van Dyk, S. et al.\ 2003, The IRSA 2MASS All-Sky Point Source Catalog, NASA/IPAC Infrared 
Science Archive

\bibitem[1997]{1997Msngr..87...27E} Epchtein, N., et al.\ 
1997, The Messenger, 87, 27 

\bibitem[1986]{1986AJ.....92.1210E} Evans, D.~S., McWilliam, 
A., Sandmann, W.~H., \& Frueh, M.\ 1986, \aj, 92, 1210 

\bibitem[2004]{2004A&A...419..285F} Fors, O., Richichi, A., 
N{\'u}{\~n}ez, J., Prades, A.\ 2004, \aap, 419, 285 

\bibitem[2006a]{SEA_2006} Fors O., Richichi, A., Mason, E., Stegmeier, J.,
Chandrasekhar, T.\ 2006, Highlights of Spanish Astrophysics IV, Figueras et al.
(Eds.), Springer 2007

\bibitem[2006b]{2006PhDT.........9F} Fors, O.\ 2006, Ph.D.~Thesis, Departament d'Astronomia i Meteorologia, Universitat de Barcelona

\bibitem[1975]{1975ascp.book.....H} Heckmann, O.\ 1975, 
Hamburg-Bergedorf: Hamburger Sternwarte, 1975, edited by Dieckvoss, W.

\bibitem[1969]{1969tmss.book.....N} Neugebauer, G., 
\& Leighton, R.~B.\ 1969, NASA SP, Washington: NASA, 1969

\bibitem[1985]{AR_thesis} Richichi, A.\ 1985, Thesis, 
Faculty of Physics, Florence University (in italian)

\bibitem[1989]{1989A&A...226..366R} Richichi, A.\ 1989, \aap, 226, 366 

\bibitem[1996]{1996AJ....112.2786R} Richichi, A., Baffa, C., Calamai,  G., Lisi, F.\ 1996, \aj, 112, 2786

\bibitem[1997]{1997IAUS..158...71R} Richichi, A.\ 1997, IAU 
Symp.~158: Very High Angular Resolution Imaging, 158, 71 

\bibitem[2002]{2002A&A...382..178R} Richichi, A., Calamai, 
G., \& Stecklum, B.\ 2002, \aap, 382, 178 

\bibitem[2002]{2002A&A...386..492R} Richichi, A., \& 
Percheron, I.\ 2002, \aap, 386, 492 

\bibitem[2006a]{2006A&A...445.1081R} Richichi, A., Fors, O., Merino, M. et al.\ 
2006, \aap, 445, 1081 

\bibitem[2006b]{2006Msngr.126....24R} Richichi, A., Fors, O., Mason, E., Stegmeier, J.\ 2006, The Messenger, 126, 24 

\bibitem[2007]{vlti_2007} Richichi, A.\ 2007, ESO Workshop  
The power of optical/IR interferometry:
recent scientific results and 2nd generation VLTI instrumentationi,
Richichi, A., Delplancke, F., Paresce, F., Chelli, A. (eds.), Springer 2007, p.31.

\bibitem[1994]{1994A&A...288..342S} Starck, J.-L., \& 
Murtagh, F.\ 1994, \aap, 288, 342

\end{thebibliography}
\end{document}